\documentclass[aps,prd,preprint,showpacs,nofootinbib,eqsecnum,amsmath,amssymb]{revtex4-1}
\usepackage[colorlinks=true,urlcolor=blue,linkcolor=blue,citecolor=blue]{hyperref}
\usepackage{amsmath} 
\usepackage{amssymb}
\usepackage{epsf}
\usepackage{color}
\usepackage{graphicx}
\usepackage{dcolumn}
\usepackage{bm}

\begin{document}

\title{Spin as a probe of axion physics in general relativity}

\author{\firstname{Yuri N.}~\surname{Obukhov}}
\email{obukhov@ibrae.ac.ru}
\affiliation{Theoretical Physics Laboratory, Nuclear Safety Institute,
Russian Academy of Sciences, B. Tulskaya 52, 115191 Moscow, Russia}
 
\begin{abstract}
The dynamics of spin in external electromagnetic, gravitational, and axion fields is analysed in the framework of the gravitoelectromagnetism approach in Einstein's general relativity theory. We consistently extend the recent studies from the flat Minkowski geometry to the curved spacetime manifolds, contributing to the discussion of the possible new role of a precessing spin as an ``axion antenna'' that can be used to detect the hypothetical axion-like dark matter. The formalism developed helps to clarify the subtle influence of the gravitational/inertial and axion fields in the ultra-sensitive high-energy spin experiments with charged particles and neutrons at accelerators and storage rings devoted to testing fundamental physical symmetries, including attempts to establish the nature of dark matter in the Universe. 
\end{abstract}

\maketitle

\section{Introduction}\label{intro}

Although the existence of an axion, a pseudoscalar particle that was originally proposed by Peccei and Quinn \cite{PQ} as a solution of the strong $CP$ problem in quantum chromodynamics, was not substantiated by the laboratory experiments and astrophysical observations, the hypothetical weakly interacting light axion-like particles are currently considered as viable candidates which could possibly explain the physical nature of the dark matter \cite{Dark1,Dark2}, and considerable experimental efforts are made to detect axions in the laboratories and to find their manifestations in astrophysics and cosmology \cite{Graham:2015,Yannis:2022,Sikivie:2021}. 

Among other research methods under development, much of attention is attracted to making use of the spin of particles in precision high-energy experiments on accelerators and storage rings as a new type of detector of axions \cite{Graham:2013,Budker:2014,Abel:2017,Chang:2019,Dror:2023}. Most recently, the concept of spin as an axion antenna was put forward \cite{Silenko:2022,Nikolaev:2022,UFN}, and the prospects of using the spin dynamics in storage rings was analysed for NMR-type experiments. In particular, since the spins in accelerators move with ultra-relativistic velocities, it was shown that one can expect a $10^3$ times amplification of the axion wind effect, as compared to fields acting on static spins.

The special-relativistic quantum-mechanical theory was developed in \cite{Silenko:2022} for the spin dynamics under the action of a pseudoscalar axion field, where the key step was the rigorous construction of the Foldy-Wouthuysen (FW) representation and derivation of the semiclassical approximation for the spin $1/2$ particle. Subsequently, the feasibility of using this approach was demonstrated \cite{Nikolaev:2022} for any storage ring equipped with internal polarimeter for the radial polarization of the beam, suggesting new possibilities of search for axions, in particular at NICA (Dubna) \cite{Senichev:2022}, COSY (Juelich) \cite{Karanth:2023}, or at planned PTR (Prototype Test Ring).

Given that our experimental laboratories are located on a gravitating and rotating Earth, it is important to carefully study the gravitational and inertial effects in the spin dynamics, in particular, since the role of these effects, at the anticipated levels of accuracy, even exceeds the role of the proton electric dipole moment (EDM), \cite{UFN}.

The purpose of the present paper is to extend the previous results \cite{Silenko:2022,Nikolaev:2022} from the flat Minkowski geometry to the curved spacetime manifolds. For all practical purposes, in experiments in terrestrial laboratories and observations in the solar system, it is sufficient to consider the linear approximation in the framework of the gravitoelectromagnetism \cite{Mashhoon:2007,Ruggiero:2002,Mashhoon:2007,Schaefer:2009,Ruggiero:2021}. The corresponding general-relativistic extension will be constructed here on the basis of the formalism \cite{OST:2016,OST:2017} developed earlier for the quantum spin dynamics in external electromagnetic, inertial and gravitational fields. 

Our main conventions and notations are the same as in Refs. \cite{OST:2016,OST:2017}. In particular, the world indices are labeled by Latin letters $i,j,k,\ldots = 0,1,2,3$ (for example, the local spacetime coordinates $x^i$ and the holonomic coframe $dx^i$), while Greek letters from the beginning of the alphabet are used for anholonomic (tetrad) indices, $\alpha,\beta,\ldots = 0,1,2,3$ (e.g., the orthonormal coframe $\vartheta^\alpha$). Furthermore, spatial indices are denoted by Latin letters from the beginning of the alphabet, $a,b,c,\ldots = 1,2,3$. In order to distinguish separate tetrad indices we put hats over them. The metric of the flat Minkowski spacetime reads $g_{\alpha\beta} = {\rm diag}(c^2, -1, -1, -1)$, where $c$ is the velocity of light, and the totally antisymmetric Levi-Civita tensor $\eta_{\alpha\beta\mu\nu}$ has the only nontrivial component $\eta_{\hat{0}\hat{1}\hat{2}\hat{3}} = c$, so that $\eta_{\hat{0}abc} = c\epsilon_{abc}$ with the three-dimensional Levi-Civita tensor $\epsilon_{abc}$. The spatial indices ($a,b,\dots = 1,2,3$) of the tensor objects are raised and lowered with the help of the Euclidean 3-dimensional metric $\delta_{ab}$. In the relativistic spinor theory, the four Dirac matrices $\gamma^\alpha$, $\alpha = 0,1,2,3$, satisfy the standard anticommutation condition $\gamma^\alpha\gamma^\beta + \gamma^\beta\gamma^\alpha = 2g^{\alpha\beta}$. As usual, $\sigma^{\alpha\beta} = {\frac i2}\left(\gamma^\alpha \gamma^\beta - \gamma^\beta\gamma^\alpha \right)$ are the generators of the local Lorentz transformations of the 4-spinor field. For the Dirac matrices as well as for the gauge-theoretic notions and objects (including electrodynamics) we use the conventions of Bogoliubov-Shirkov \cite{BS}.

\section{Gravitoelectromagnetism and axion electrodynamics}\label{GEM}

In the linear perturbation approach to gravitoelectromagnetism \cite{Mashhoon:2007,Ruggiero:2002,Mashhoon:2007,Schaefer:2009,Ruggiero:2021}, the spacetime interval is described in terms of the gravitoelectric ${\mathit\Phi}$ and the gravitomagnetic $\bm{\mathcal A}$ potentials as
\begin{equation}\label{dsGEM}
ds^2 = \left(1 - {\frac {\mathit \Phi}{c^2}}\right)^2c^2dt^2 + {\frac 4c}(\bm{\mathcal A}\cdot d\bm{x})dt
- \left(1 + {\frac {\mathit \Phi}{c^2}}\right)^2d\bm{x}\cdot d\bm{x}.
\end{equation}
When the gravitational field is created by a body with the mass $M$ and the angular momentum $\bm{J}$, the gravitoelectromagnetic fields far from the massive source are given by the configuration that was first derived by Lense and Thirring \cite{Lense:1918,Hehl:1984}:
\begin{equation}\label{geLT}
{\mathit \Phi} = {\frac {GM}{r}},\qquad \bm{\mathcal A} = {\frac {G\,\bm{J}\times\bm{r}}{c\,r^3}}.
\end{equation}
Here $G$ is Newton's gravitational constant. Note that we use the calligraphic font for the gravitoelectromagnetic potentials $({\mathit\Phi}, \bm{\mathcal A})$ to distinguish them from the electromagnetic potentials $A_i = (\Phi, \bm{A})$. 

Before we turn to fermions, it is instructive to consider electrodynamics. Maxwell's equations on curved spacetime \cite{YNO:2021} have the form of the electrodynamics in a medium 
\begin{eqnarray}\label{max1}
\bm{\nabla}\times \bm{E} + {\frac {\partial\bm{B}}{\partial t}} &=& 0,
\qquad \bm{\nabla}\cdot\bm{B} = 0,\\
\bm{\nabla}\times \bm{H} - {\frac {\partial\bm{D}}{\partial t}} &=& \bm{J}^{\rm e},
\qquad \bm{\nabla}\cdot\bm{D} = \rho^{\rm e}.\label{max2}
\end{eqnarray}
The matter sources are the electric current density $\bm{J}^{\rm e}$ and the electric charge density $\rho^{\rm e}$. The influence of the inertia and gravity is encoded in the Maxwell-Lorentz constitutive relation between the electric and magnetic fields $\bm{E}, \bm{B}$ and the electric and magnetic excitations $\bm{D}, \bm{H}$.

The pseudoscalar axion field $\varphi$ couples to electromagnetic field in a very peculiar way, being the isotropic irreducible magnetoelectric part of the general linear constitutive tensor \cite{Birk}. Specializing to the GEM geometry (\ref{dsGEM}), the axion electrodynamics then arises as the constitutive relation
\begin{align}
\bm{D} &= \varepsilon_0\varepsilon_g\,\bm{E} + {\frac {\lambda_0}{c^2}}\,\bm{\mathcal A}\times\bm{B}
+ \varphi\,\bm{B},\label{DE}\\
\bm{H} &= {\frac 1{\mu_0\mu_g}}\,\bm{B} + {\frac {\lambda_0}{c^2}}\,\bm{\mathcal A}\times\bm{E}
- \varphi\,\bm{E}.\label{HB}
\end{align}
Here $\varepsilon_0$ and $\mu_0$ are the electric and magnetic constants of the vacuum, $\lambda_0 = \sqrt{\varepsilon_0/\mu_0}$ (it is worthwhile to note that the velocity of light $c = 1/\sqrt{\varepsilon_0\mu_0}$). The properties of this ``medium'' are determined by the gravitational field: the gravitoelectric potential ${\mathit \Phi}$ introduces the effective permittivity and permeability
\begin{equation}\label{mug}
\varepsilon_g = \mu_g = \left(1 + {\frac {\mathit \Phi}{c^2}}\right)^2,
\end{equation}
whereas the gravitomagnetic potential $\bm{\mathcal A}$ is responsible for the effective magnetoelectric effects.

Apparently, the first who discussed the axion electrodynamics in the gravitational context were Schr\"odinger \cite{Schroedinger} (see p. 25) and Dicke \cite{Dicke} (see Appendix 4 on p. 47).

\section{Fermion particle in external fields}\label{dirac}

The dynamics of a fermion particle with the spin $1/2$, the rest mass $m$ and electric charge $q$ is derived from the Lagrangian 
\begin{eqnarray}
L &=& {\frac {i\hbar}{2}}\left(\overline{\Psi}\gamma^\alpha D_\alpha\Psi
- D_\alpha\overline{\Psi}\gamma^\alpha\Psi\right) - mc\,\overline{\Psi}\Psi \nonumber\\
&& +\,{\frac{\mu'}{2c}}\overline{\Psi}\sigma^{\alpha\beta}\Psi F_{\alpha\beta} 
+ {\frac{\delta'}{2}}\overline{\Psi}\sigma^{\alpha\beta}\Psi\widetilde{F}{}_{\alpha\beta} 
-\,{\frac{\hbar\,g_f}{2f_{(a)}}}\,\overline{\Psi}\gamma^\alpha\gamma_5\Psi
\left(e_\alpha^i\partial_i\varphi\right).\label{LD}
\end{eqnarray}
The first two terms describe a Dirac wave function $\Psi$ minimally coupled to the electromagnetic field $A_i$ and the gravitational field $(e_i^\alpha, \Gamma_i{}^{\beta\gamma})$ potentials, which is encoded in the covariant spinor derivative
\begin{equation}
D_\alpha = e_\alpha^i\Bigl(\partial _i - {\frac {iq}{\hbar}}\,A_i
+ {\frac i4}\Gamma_i{}^{\beta\gamma}\sigma_{\beta\gamma}\Bigr).\label{eqin2}
\end{equation}
The next two Pauli terms describe the non-minimal coupling of the electromagnetic field strength $F_{\alpha\beta}$ and its dual $\widetilde{F}{}_{\alpha\beta} = {\frac 12}\eta_{\alpha\beta\mu\nu}F^{\mu\nu}$ to an anomalous magnetic moment (AMM) and an electric dipole moment (EDM) of the fermion. The respective coupling parameters have the dimension $[\mu'] = [q\hbar/2m]$ of the magnetic dipole (nuclear magneton), and $[\delta'] = [q\,l]$ of the electric dipole (charge times length). The anomalous magnetic moment is usually given in terms of the magneton $\mu_0 = {\frac {q\hbar}{2m}}$, and in a similar way one can introduce a convenient unit of an electric dipole moment. A reasonable definition is the electric charge times the electron Compton length: $\delta_0 = q\,{\frac {\hbar}{mc}}$. Then we have for both types of the dipole moments:
\begin{equation}
\mu' = a\,{\frac {q\hbar}{2m}},\qquad \delta' = b\,{\frac {q\hbar}{2mc}}.\label{mude}
\end{equation}
The two dimensionless constant parameters $a = (g -2)/2$ (with $g$ as the gyromagnetic factor) and $b$ characterize the magnitude of the anomalous magnetic and electric dipole moments, respectively.

The last term in (\ref{LD}) manifests the interaction of the fermion with the pseudoscalar axion field $\varphi$, where $g_f \sim 1$ is a model-dependent dimensionless constant, and the coupling constant $f_{(a)}$ (with the dimension of $\varphi$) is related to the mass $m_{(a)}$ of the axion via $f_{(a)}m_{(a)} \approx f_\pi m_\pi \sqrt{m_u m_d}/(m_u+m_d)$, with quark masses $m_{u,d}$, and the mass $m_\pi$ and decay constant $f_\pi$ of the pion \cite{Weinberg:1978}.

Making use of the Schwinger coframe $\vartheta^\alpha = e^\alpha_idx^i$, 
\begin{equation}\label{cof}
\vartheta^{\hat 0} = \left(1 - {\frac {\mathit \Phi}{c^2}}\right)dt,\qquad
\vartheta^{\hat a} = \left(1 + {\frac {\mathit \Phi}{c^2}}\right)dx^a - {\frac 2c}{\mathcal A}^adt,
\end{equation}
the Dirac equation derived from (\ref{LD}) can be recast into the Schr\"odinger form
\begin{equation}\label{sch}
i\hbar\frac{\partial \psi} {\partial t}= {\cal H}\psi
\end{equation}
with the Hermitian Hamiltonian
\begin{eqnarray}
{\cal H} &=& {\cal H}^{\rm GEM} + {\cal H}^{\rm ax},\label{H0}\\
{\cal H}^{\rm GEM} &=& mc^2\beta^g + q\Phi + {\frac c 2}\left(\bm{\pi}\cdot\bm{\alpha}^g + \bm{\alpha}^g
\cdot\bm{\pi}\right)\nonumber\\
&& + \,{\frac {\hbar}{2c}}\,\bm{\Sigma}\cdot(\bm{\nabla}\times\bm{\mathcal A}) 
- \beta^g\left(\bm{\Sigma}\cdot\bm{\mathcal M} + i\bm{\alpha}\cdot\bm{\mathcal P}\right),\label{Hgem}\\
{\cal H}^{\rm ax} &=& {\frac \hbar 2}\,{\frac {g_f}{f_{(a)}}}\left[{\frac {c}{\mu_g}}\bm{\Sigma}\cdot
\bm{\nabla}\varphi - \gamma_5\,\Bigl(\partial_t\varphi + {\frac 2c}\bm{\mathcal A}\cdot
\bm{\nabla}\varphi\Bigr)\right].\label{Hax}
\end{eqnarray}
The Hamiltonian ${\cal H}^{\rm GEM}$ determines the quantum dynamics of a fermion particle under the action of the electromagnetic field $(\Phi, \bm{A})$ (with $\bm\pi=-i\hbar\bm{\nabla} - q\bm A$ as the kinetic momentum operator) and the gravitoelectromagnetic field $({\mathit\Phi}, \bm{\mathcal A})$. The effects of the latter are conveniently described by making use of the condensed notation
\begin{equation}\label{abe}
\beta^g := {\frac {\beta}{1 + {\frac {\mathit \Phi}{c^2}}}},\qquad
\bm{\alpha}^g := {\frac {\bm{\alpha}}{\mu_g}} + {\frac 2{c^2}}\bm{\mathcal A}.
\end{equation}
The $4\times 4$ Dirac matrices have their usual form here, namely \cite{BS}:
\begin{equation}
\beta = \Biggl(\begin{array}{cc}I_2 & 0\\ 0 & - I_2\end{array}\Biggr),\quad \bm{\alpha} =
\Biggl(\begin{array}{cc}0 & \bm{\sigma}\\ \bm{\sigma} & 0\end{array}\Biggr),\quad \bm{\Sigma}
= \Biggl(\begin{array}{cc}\bm{\sigma} & 0\\ 0 & \bm{\sigma}\end{array}\Biggr),\quad
\gamma_5 = \Biggl(\begin{array}{cc}0 & -I_2\\ -I_2 & 0\end{array}\Biggr),
\end{equation}
where $I_2$ is the $2\times 2$ unit matrix, and $\bm{\sigma}$ are the Pauli matrices. 
The last term in (\ref{Hgem}) accounts for the nonminimal electromagnetic coupling in terms of the generalized polarization and magnetization (both with dimension of energy)
\begin{align}\label{Ma}
\bm{\mathcal M} = \mu'\bm{\mathfrak B} + \delta'\bm{\mathfrak E} &=
{\frac {\hbar q}{2m}}\Bigl(a\,\bm{\mathfrak B} + {\frac bc}\,\bm{\mathfrak E}\Bigr),\\
\bm{\mathcal P} = c\delta'\bm{\mathfrak B} - {\frac {\mu'}c}\,\bm{\mathfrak E} &=
{\frac {\hbar q}{2m}}\left(b\,\bm{\mathfrak B} - {\frac ac}\,\bm{\mathfrak E}\right),\label{Pa}
\end{align}
constructed from the anholonomic electric and magnetic fields which in the coframe (\ref{cof}) read
\begin{equation}\label{EB}
\bm{\mathfrak{E}} = \bm{E} + {\frac 2c}\bm{\mathcal A}\times\bm{B},\qquad
\bm{\mathfrak{B}} = {\frac 1{\mu_g}}\,\bm{B},
\end{equation}
where $\bm{E} = -\,\nabla\Phi-\frac{\partial\bm A}{\partial t}$ and $\bm{B}=\nabla\times\bm A$. We use different fonts (Gothic and Roman) to distinguish the anholonomic and holonomic components, respectively.

\section{Quantum spin dynamics}\label{spin}

To determine the physical content of the Schr\"{o}dinger equation (\ref{sch}), one should pass to the Foldy-Wouthuysen (FW) representation. One can construct the FW transformation for the Dirac Hamiltonian (\ref{H0})-(\ref{Hax}) with the general method developed in Refs. \cite{OST:2016,OST:2017}. 

For the practical problems in the high-energy particle physics in accelerators and storage rings, it is sufficient to work with quasiclassical quantities and equations. After a lengthy computation along the lines of \cite{OST:2016,OST:2017}, we find the nonrelativistic FW Hamiltonian
\begin{equation}\label{Hamlt}
{\cal H}_{FW} = {\frac 1{1 + {\frac {\mathit\Phi}{c^2}}}}\Bigl(mc^2 + {\frac 1{2m}}\bm{P}^2\Bigr) + q\Phi
+ {\frac \hbar2}\bm{\sigma}\cdot\bm{\Omega},\qquad \bm{P} = \bm{\pi} + 2m\bm{\mathcal A}/c,
\end{equation}
which then describes the precession of the mean spin 3-vector ${\bm s}$:
\begin{eqnarray}
{\frac {d{\bm s}}{dt}} = \bm{\Omega}\times{\bm s}.\label{dots}
\end{eqnarray}
The precession angular velocity is explicitly given by the sum of four terms:
\begin{equation}\label{Omega}
\bm{\Omega} = \bm{\Omega}^{\rm em} + \bm{\Omega}^{\rm dip} + \bm{\Omega}^{\rm GEM} + \bm{\Omega}^{\rm ax},
\end{equation}
where the first two contributions account for the electromagnetic effects,
\begin{align}\label{OmegaE}
\bm{\Omega}^{\rm em} &= {\frac {q}{m}}\left[- \,{\frac 1{\gamma}}\,\bm{\mathfrak B} +
{\frac {1}{\gamma + 1}}{\frac {\widehat{\bm v}\times\bm{\mathfrak E}}{c^2}}\right],\\
\bm{\Omega}^{\rm dip} &= {\frac {q}{m}}\left[- \,{\frac 1{\gamma}}\,\bm{\mathfrak B}^{\rm dip} + 
{\frac {1}{\gamma + 1}}{\frac {\widehat{\bm v}\times\bm{\mathfrak E}^{\rm dip}}{c^2}}\right],\label{OmegaD}
\end{align}
and the two last terms in (\ref{Omega}) are due to influence of non-electromagnetic fields
\begin{align}\label{OmegaG}
\bm{\Omega}^{\rm GEM} &= - \,{\frac 1{\gamma}}\,\bm{\mathcal B}^{g} +
{\frac {1}{\gamma + 1}}{\frac {\widehat{\bm v}\times\bm{\mathcal E}^g}{c^2}},\\
\bm{\Omega}^{\rm ax} &= -\,{\frac 1{\gamma}}\,\bm{\mathcal B}^{\rm ax} + 
{\frac {1}{\gamma + 1}}{\frac {\widehat{\bm v}\times\bm{\mathcal E}^{\rm ax}}{c^2}}.\label{OmegaA}
\end{align}
The Lorentz factor $\gamma = 1/\sqrt{1 - \widehat{v}{}^2/c^2}$ is constructed from particle's velocity $\widehat{\bm{v}}$ evaluated with respect to the coframe (\ref{cof}).

By analogy with the true magnetic $\bm{\mathfrak B}$ and electric $\bm{\mathfrak E}$ fields, it is natural to view the vectors that enter (\ref{OmegaD})-(\ref{OmegaA}) as {\it effective} magnetic and electric fields. There are three types of them: the dipole-magnetic and the dipole-electric fields
\begin{align}
\bm{\mathfrak B}^{\rm dip} =&\, \gamma\Bigl[ a\Bigl(\bm{\mathfrak B} - {\frac {\widehat{\bm{v}}\times
\bm{\mathfrak E}}{c^2}} - {\frac{\widehat{\bm{v}}\,(\widehat{\bm{v}}\cdot\bm{\mathfrak B})}{c^2}}\Bigr)\nonumber\\
& + {\frac bc}\Bigl(\bm{\mathfrak E} + \widehat{\bm{v}}\times\bm{\mathfrak B}
- {\frac {\widehat{\bm{v}}\,(\widehat{\bm{v}}\cdot\bm{\mathfrak E})}{c^2}}\Bigr)\Bigr],\label{Bdip}\\
\bm{\mathfrak E}^{\rm dip} =&\, -\,\gamma\widehat{\bm{v}}\times\Bigl[a\Bigl(\bm{\mathfrak B} 
- {\frac {\widehat{\bm{v}}\times\bm{\mathfrak E}}{c^2}}\Bigr) + {\frac bc}\Bigl(\bm{\mathfrak E}
+ \widehat{\bm{v}}\times\bm{\mathfrak B}\Bigr)\Bigr],\label{Edip}
\end{align}
the gravito-magnetic and the gravito-electric fields
\begin{align}\label{Bgem}
\bm{\mathcal B}^g &= -\,{\frac {\gamma}{c}}\,\bm{\nabla}\times \bm{\mathcal A}
- {\frac {\gamma}{c^2}}\,\widehat{\bm{v}}\times\bm{\nabla}{\mathit\Phi},\\
\bm{\mathcal E}^g &= \gamma\,\bm{\nabla}{\mathit\Phi},\label{Egem}
\end{align}
and, finally, the axi-magnetic and the axi-electric fields,
\begin{align}\label{Bax}
\bm{\mathcal B}^{\rm ax} &= -\,{\frac {g_f}{f_{(a)}}}{\frac {\gamma}{\left(1 + {\frac {\mathit \Phi}{c^2}}
\right)}}\,\left[c\,\bm{\nabla}\varphi + {\frac {\widehat{\bm{v}}}{c}}\mu_g\,\Bigl(\partial_t\varphi
+ {\frac 2c}\bm{\mathcal A}\cdot\bm{\nabla}\varphi\Bigr)\right],\\
\bm{\mathcal E}^{\rm ax} &= {\frac {g_f}{f_{(a)}}}{\frac {\gamma c}{\left(1 + {\frac {\mathit \Phi}{c^2}}
\right)}}\,\widehat{\bm{v}}\times\bm{\nabla}\varphi.\label{Eax}
\end{align}
Substituting (\ref{Bdip})-(\ref{Eax}) into (\ref{OmegaD})-(\ref{OmegaA}), we find explicitly
\begin{align}
\bm{\Omega}^{\rm dip} =&\,-\,{\frac qm}\Bigl\{\Bigl[ a\Bigl(\bm{\mathfrak B}
- {\frac {\widehat{\bm{v}}\times\bm{\mathfrak E}}{c^2}} - {\frac {\gamma}{\gamma + 1}}
\,{\frac{\widehat{\bm{v}}\,(\widehat{\bm{v}}\cdot\bm{\mathfrak B})}{c^2}}\Bigr)\nonumber\\
&\qquad + {\frac bc}\Bigl(\bm{\mathfrak E} + \widehat{\bm{v}}\times\bm{\mathfrak B}
- {\frac {\gamma}{\gamma + 1}}\,{\frac {\widehat{\bm{v}}\,(\widehat{\bm{v}}\cdot
\bm{\mathfrak E})}{c^2}}\Bigr)\Bigr]\Bigr\},\label{OmD}\\
\bm{\Omega}^{\rm GEM} =&\, {\frac {1}{c}}\,\bm{\nabla}\times \bm{\mathcal A} + {\frac {(2\gamma + 1)}
{(\gamma + 1)c^2}}\,\widehat{\bm{v}}\times\bm{\nabla}{\mathit\Phi},\label{OmG}\\
\bm{\Omega}^{\rm ax} =&\, {\frac {g_f}{f_{(a)}}}{\frac {1}{\left(1 + {\frac {\mathit \Phi}{c^2}}
\right)}}\,\left\{{\frac c\gamma}\,\bm{\nabla}\varphi + {\frac {\widehat{\bm{v}}}{c}}\Bigl[\mu_g\,\Bigl(
\partial_t\varphi + {\frac 2c}\bm{\mathcal A}\cdot\bm{\nabla}\varphi\Bigr) + 
{\frac {\gamma}{\gamma + 1}}\,\widehat{\bm{v}}\cdot\bm{\nabla}\varphi\Bigr]\right\}.\label{OmA}
\end{align}
Curiously enough, the precession velocity (\ref{OmD}) due to the anomalous magnetic dipole $a$ and electric dipole $b$ moments is determined by the Lorentz-transformed magnetic and electric fields.

In the special case of a rotating massive body, we substitute (\ref{geLT}) into (\ref{OmG}) to recast the spin precession in the gravitational field into a sum of the two terms which are known as the ``de Sitter precession'' (or the geodetic precession) and the ``Lense-Thirring precession'', respectively, $\bm{\Omega}^{\rm GEM} = \bm{\Omega}^{\rm dS} + \bm{\Omega}^{\rm LT}$:
\begin{eqnarray}\label{Osum}
\bm{\Omega}^{\rm dS} &=& {\frac {(2\gamma + 1)}{(\gamma + 1)}}\,
{\frac {GM\,\bm{r}\times\widehat{\bm{v}}}{c^2\,r^3}},\label{omDS}\\
\bm{\Omega}^{\rm LT} &=& {\frac {G}{c^2\,r^3}}\left[
{\frac {3(\bm{J}\cdot\bm{r})\,\bm{r}}{r^2}} - \bm{J}\right].\label{omLT}
\end{eqnarray}
The validity of this result for the gravitational field of the Earth was confirmed in the Gravity Probe B space experiment \cite{GPB,Everitt:2015}.

Although the dynamics of spin is complicated, and all four contributions (\ref{OmegaE})-(\ref{OmegaA}) are important, in general, the most interesting is the new result (\ref{OmA}) which extends the studies of spin as antenna for the axion \cite{Silenko:2022,Nikolaev:2022} from the flat space to the curved geometries (\ref{dsGEM}) in the framework of the GEM approach. 

We confirm here the conclusions derived earlier for the flat spacetime, and find corrections due to the gravitational and inertial fields. It is worthwhile to notice the peculiar ``mixing'' of axion effects with inertial/gravitational ones. In particular, under the conditions of the high-energy experiments in accelerators located on the Earth that rotates with the angular velocity $\bm{\omega}_\oplus$, one can approximately take $\mathit{\Phi} = 1$ and ${\mathcal A} = -\,c\,\bm{\omega}_\oplus\times\bm{r}/2$ (so that $\bm{\mathcal B}^g = \bm{\omega}_\oplus$), and then the axion contribution in the nonrelativistic Hamiltonian (\ref{Hamlt}) reads
\begin{eqnarray}
{\mathcal H}{}_{FW}^{\rm ax} = -\,{\frac {\hbar}{2}}\,\bm{\sigma}\cdot\bm{\mathcal B}^{\rm ax}
= {\frac {\hbar cg_f}{2f_{(a)}}}\,\bm{\sigma}\cdot\Bigl[\bm{\nabla}\varphi
+ {\frac {\bm{p}}{mc^2}}\,{\frac {d\varphi}{dt}}\Bigr]\,.\label{axiH}
\end{eqnarray}
This agrees with the flat space results \cite{Silenko:2022,Nikolaev:2022}, however, the new feature is that the rate of the change of the axion field is given by the material derivative ${\frac {d\varphi}{dt}} = {\frac {\partial\varphi}{\partial t}} + \bm{v}^{\rm rot}\cdot \bm{\nabla}\varphi$, where $\bm{v}^{\rm rot} = \bm{\omega}_\oplus\times\bm{r}$ is the dragging velocity due to the motion of the frame, located on the rotating Earth. In this way, a longitudinal pseudomagnetic field acting on a spin can be generated not only by a time-dependent axion configuration, but also by a static inhomogeneous axion field.

Note that in addition to the direct influence of the Earth's gravity and rotation through the spacetime GEM metric (\ref{dsGEM}) and the coframe components (\ref{cof}) in the structure of the gravito-magnetic (\ref{Bgem}) and gravito-electric (\ref{Egem}) fields, the gravitational field implicitly manifests its influence also through the form of the axion field obtained as a solution of the scalar wave equation in the curved spacetime. The corresponding analysis of such effects was carried out in \cite{Stadnik:2014}, however, without taking into account the Earth's rotation. 

For completeness, we have to mention that our results also include another possible interaction mechanism of a pseudoscalar axion field with particle's spin via the EDM Pauli term ${\frac{\delta'}{2}}\overline{\Psi}\sigma^{\alpha\beta}\Psi\widetilde{F}{}_{\alpha\beta}$. Technically, this amounts to the shift of the electric dipole parameter $b = b_0 + \kappa_d\varphi/f_{(a)}$, where $b_0$ accounts for the constant EDM, and the dimensionless model-dependent factor $\kappa_d \approx 10^{-2}$, \cite{UFN}. For the classical axion field $\varphi = \varphi_0\cos(\omega_{(a)} - \bm{k}_{(a)}\cdot\bm{x})$ in the invisible halo of our Galaxy, this produces an oscillating contribution in the precession angular velocity (\ref{OmD}).

\section{Classical spin dynamics}\label{BMT}

The classical theory of spin, which was originally put forward by Frenkel and Thomas, and further developed by Mathisson, Papapetrou and Dixon, gives an adequate description of a particle with spin, and it underlies the analysis of the dynamics of polarized particles in accelerators and storage rings \cite{BMT1,BMT2,BMT3,BMT4,BMT5}, for a detailed overview and more references see \cite{UFN}. 

In the framework of the classical approach, the motion of a test spinning particle is characterized by the 4-velocity $U^\alpha$ and the 4-vector of spin $S^\alpha$, which satisfy the normalization $U_{\alpha} U^\alpha = c^2$ and the orthogonality condition $S_{\alpha} U^\alpha = 0$. The motion of a relativistic particle with mass $m$, electric charge $q$ and anomalous dipole moments $\mu', \delta'$ in the electromagnetic, gravitational and axion fields, is described by the dynamical system 
\begin{eqnarray}
{\frac {DU^\alpha}{d\tau}} &=& {\frac {dU^\alpha}{d\tau}} + U^i\Gamma_{i\beta}{}^\alpha U^\beta
= -\,{\frac qm}\,g^{\alpha\beta}F_{\beta\gamma}U^\gamma,\label{DUG}\\
{\frac {DS^\alpha}{d\tau}} &=& {\frac {dS^\alpha}{d\tau}} + U^i\Gamma_{i\beta} {}^\alpha S^\beta
=  -\,{\frac qm}\,g^{\alpha\beta}F_{\beta\gamma}S^\gamma + {\frac{g_f}{f_{(a)}}}\,\eta^{\alpha\beta\gamma\delta}U_\delta
\,(e^i_\gamma\,\partial_i\varphi)\,S_\beta\nonumber\\
&& -\,{\frac 2\hbar}\left[M^\alpha{}_\beta + {\frac {1}{c^2}}\left(M_{\beta\gamma}U^\alpha
U^\gamma - M^{\alpha\gamma}U_\beta U_\gamma\right)\right]S^\beta.\label{DSG}
\end{eqnarray}
Here, particle's trajectory is parametrized by the proper time $\tau$, and we introduced the polarization tensor 
\begin{equation}
M_{\alpha\beta} = \mu' F_{\alpha\beta} + c\delta'\,\widetilde{F}_{\alpha\beta},\label{Mab}
\end{equation}
whose components, $M_{\hat{0}\hat{a}} = c{\mathcal P}_a$ and $M_{\hat{a}\hat{b}} = \epsilon_{abc}{\mathcal M}^c$, are identified with the 3-vectors $\bm{\mathcal M}$ and $\bm{\mathcal P}$ which we defined in (\ref{Ma}) and (\ref{Pa}).

Recalling that the physical spin, as an ``internal angular momentum'', is defined in particle's rest frame, we need to perform the local Lorentz transformation $U^\alpha = \Lambda^\alpha{}_\beta u^\beta$, $S^\alpha = \Lambda^\alpha{}_\beta s^\beta$ from the laboratory reference system to the rest frame in which $u^\alpha = \delta^\alpha_0 = (1, \bm{0})$ and $s^\alpha = (0, \bm{s})$. Applying this transformation to the system (\ref{DUG})-(\ref{DSG}), we recover the precession equation (\ref{dots}) for the physical spin $\bm{s}$ with the angular velocity (\ref{Omega})-(\ref{OmegaA}), which demonstrates the full agreement between the quantum-mechanical and classical dynamics of spin. Note that, technically, it is necessary to use the relation between the derivatives with respect to the proper and coordinate time, ${\frac {d}{d\tau}} = \gamma\left(1 + {\frac {\mathit \Phi}{c^2}}\right){\frac {d} {dt}}$. For an alternative discussion of the classical spin dynamics in an axion field, see \cite{Balakin:2016,Dvornikov:2019}.

\section{Discussion and conclusions}\label{conc}

In this paper, the dynamics of spin in external electromagnetic, gravitational, and axion fields is analysed in the framework of the linear GEM approach in Einstein's general relativity theory. The recent studies \cite{Silenko:2022,Nikolaev:2022} are here consistently extended from the flat Minkowski geometry to the curved spacetime manifolds. The results obtained can be directly applied to the careful account of the gravitational and inertial effects in the precision high-energy experiments at accelerators and storage rings devoted to testing fundamental physical symmetries, including attempts to establish the nature of dark matter in the Universe. Our work thus contributes to the discussion of the possible new role of a precessing spin as a detector of hypothetical axion-like dark matter.

\begin{acknowledgments}
I would like to thank Nikolai Nikolaev (Landau Institute for Theoretical Physics, Chernogolovka) for constant support and strong encouragement, and the numerous fruitful discussions with Alexander Silenko and Oleg Teryaev (JINR, Dubna) are gratefully acknowledged.  
\end{acknowledgments}

\end{document}